# Clock Tree Generation by Abutment in Synchoros VLSI Design

Dimitrios Stathis[*1], Panagiotis Chaourani[*], Syed M. A. H. Jafri[*], Ahmed Hemani[*]
[*]KTH Royal Institute of Technology, Stockholm, Sweden (e-mail: stathis, pancha, jafri, hemani@kth.se)

*Abstract*— **Synchoros VLSI design style has been proposed as an alternative to standard cell-based design. Standard cells are replaced by synchoros, large grain, VLSI design objects called SiLago (Silicon Lego) blocks. This new design style eliminates the need to synthesise ad hoc wires of any type: functional and infrastructural. SiLago blocks are organised into region instances. In a region instance, communication among SiLago blocks is synchronous and happens over a regional network on chip (NoC), whose fragments are also absorbed into SiLago blocks. Consequently, the regional NoCs get created by the abutment of SiLago blocks. The clock tree used in a region is called regional clock tree (RCT). The synchoros VLSI design style requires that the RCT, like the regional NoCs, is also created by abutting its fragments. The RCT fragments are absorbed within the SiLago blocks. The RCT created by abutment is not an ad-hoc clock tree but a structured and predictable design with known cost metrics. The design of such an RCT is the focus of this paper. The scheme is scalable, and we demonstrate that the proposed RCT can be generated for valid VLSI designs of ~1.5 million gates. The RCT created by abutment is correct by construction, and its properties are predictable. We have validated the generated RCTs with static timing analysis to validate the correct-by-construction claim. Finally, we show that the cost metrics of the SiLago RCT is comparable to the one generated by commercial EDA tools.**

## I. INTRODUCTION

This paper presents a clock tree generation scheme for a novel synchoros very large scale integration **(VLSI)** design style. The term synchoros is derived from the Greek word for space – χώρος (khôros). Synchoricity is analogous to synchronicity[2]. In synchronicity, time is uniformly discretised with clock ticks to enable the composition of synchronous digital systems in terms of component synchronous digital designs. Correspondingly, in synchoricity, space is uniformly discretised with a virtual grid to simplify the spatial and electrical composition of larger VLSI systems' in terms of component synchoros building blocks – the SiLago (Silicon Lego) blocks.

The synchoros VLSI design style is inspired by the Lego bricks. All Lego bricks are integer multiples of some standard pitch. The studs and their corresponding female receptors are also standardised and uniformly distributed. This uniform discretisation of space enables the creation of an infinite variation of Lego systems from finite types of Lego bricks, see Figure 1a.

Another inspirational example is the modern practice of building large structures from different types of prefabricated wall segments **(PFWS)**, see Figure 1c. The PWFS come in standardised dimensions, with their height and width being integer multiples of some standard pitch. They absorb fragments of plumbing, electrical, ventilation, telecom constructs, etc. As a result, larger walls are built by abutment of PFWS without needing any plumbers, carpenters, electricians etc. The convenience is great, but a more profound benefit is that all the properties of the emerging plumbing, electrical constructs etc., are accurately predictable by simply knowing the wall segments involved and their topological relationship.

Inspired by these examples, the synchoros VLSI design style, see Figure 1b, enables a) the creation of an infinite variety of complex VLSI structures from finite types of SiLago blocks. The synchoros VLSI design style makes it possible for the end-user to create a VLSI design without any need to do logic and physical synthesis, and b) the emerging VLSI designs created by abutment are not only a valid (timing and DRC[3] clean) design, but it is also predictable with the accuracy of post-layout data [1]. This paper focuses on how one category of wires – the regional clock tree **(RCT)** – is absorbed inside the SiLago blocks, and a valid and predictable RCT gets created by abutment.

In the state-of-the-art, standard-cells based design style, the cells themselves are pre-designed, but their placement and wires needed to connect, clock, power, and reset them must be synthesised anew in an ad-hoc fashion. This ad hoc synthesis of wires makes the cost metrics unpredictable for synthesis from higher abstractions. In the synchoros VLSI design style, all wires – functional and infrastructural – are absorbed in the SiLago blocks. The SiLago blocks replace standard cells as atomic building blocks. All inter-SiLago block wires are bought to the periphery at the right place and at the right metal layer to enable composition by abutment of valid neighbours.

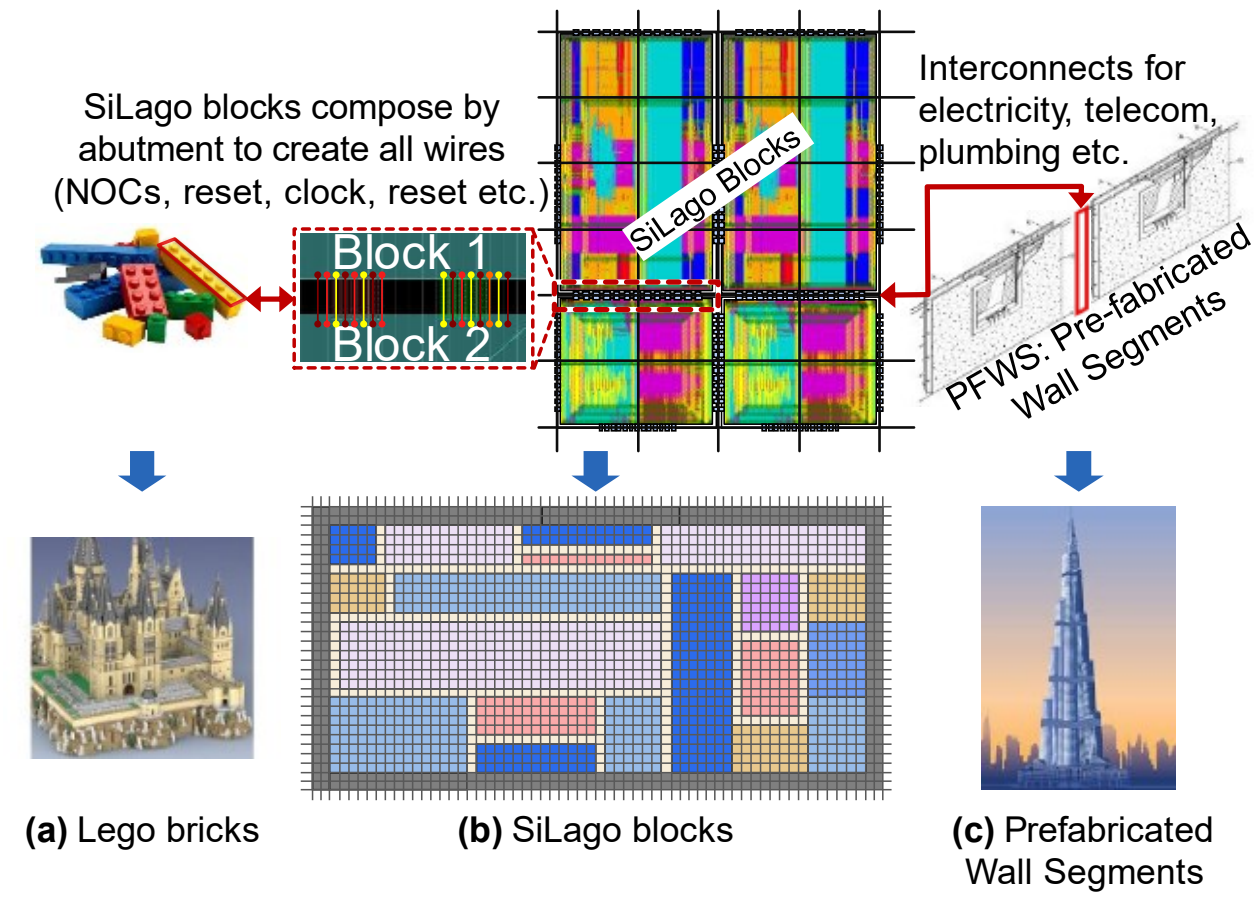

Figure 1. Synchoros VLSI Design style is inspired by Lego and prefabricated wall segments.

[1] Corresponding author
[2] Synchoricity is a different concept from synchronicity, and not a typo. We highlight the difference between the two words by underlining the n in synchronicity.
[3] Design rule check

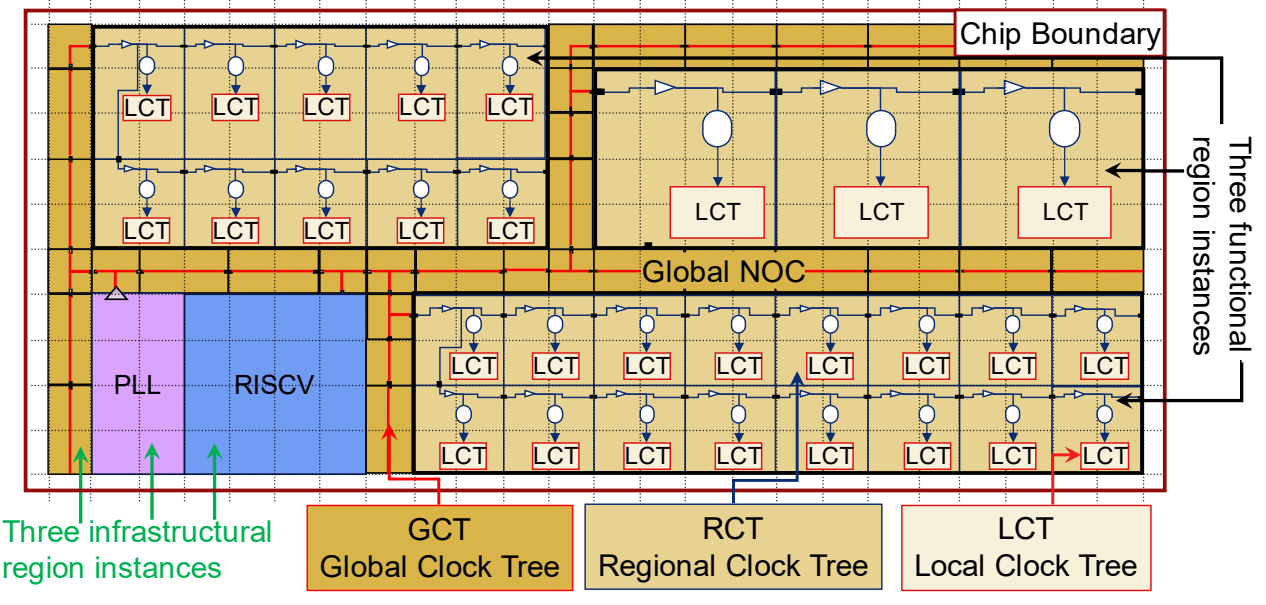

Figure 2. Example of synchoros VLSI design instance

In the synchoros VLSI design style, there are three levels of hierarchical objects (Figure 2): Local, Regional and Global.

<u>Local:</u> The SiLago blocks are at the lowest level of the hierarchy. All wires in SiLago blocks, including the Local Clock Tree **(LCT)**, are ad-hoc and synthesised by the EDA tools.

<u>Regions:</u> Regions are composed of SiLago blocks of the same type, as shown in Figure 2. Inter-SiLago block communication in a region happens over a regional Network on Chip **(NoC)**, whose wires are also absorbed within the SiLago blocks. The RCT drives the LCTs in each SiLago block. The RCT fragments and ancillary circuit to maintain the slew and minimise the skew are also absorbed in each SiLago block. As stated, this is the focus of this paper. The RCT must maintain synchronicity amongst SiLago blocks in a region, as shown in Figure 7. The figure illustrates an inter-SiLago block timing path created by a regional NoC clocked by two local clocks.

<u>Global:</u> Region instances communicate over Global NoCs, also composed of global SiLago blocks [2], see Figure 2. Global NoCs have their own Global Clock Tree **(GCT)**. The global NoCs and the GCT are not part of this paper.

The paper is organized as follows. Section II introduces in more detail the synchoros VLSI design style. It also motivates the development of synchoros VLSI design style and what benefits it brings to electronic design automation **(EDA)**. Section III presents the proposed design of the regional clock tree **(RCT)** architecture, which is the main contribution of this work. Section IV introduces the problem of RCT configuration, formulates it as an optimization problem and proposes two cost functions that can be used for the optimization. Additionally, we introduce some heuristics that can significantly reduce the solution space and the complexity of the problem. Section V presents the experimental setup used to validate the RCT, describes the experiments, and analyses their results. The proposed RCT design is compared with one generated from the state-of-the-art EDA tools. In section VI, we explore the state-of-the-art, and we reason why the state-of-the-art solutions do not satisfy the requirements of the synchoros VLSI design. Finally, in section VII, the paper is concluded with a summary of the contributions and what we believe are the limitations of this work, as well as a road map for the future.

## II. An Overview of the Synchoros VLSI Design

The essence of synchoros VLSI design was introduced as the background of this work in section I. This section presents arguments for why and how synchoros VLSI design enables design automation from higher abstractions and why this has not been possible with standard cell based design flows. Such automation aims to make ASIC comparable custom chip design accessible to all. Chip design today requires >300 million US dollars [3], 90% of which is engineering cost. We next explain the root cause of this large engineering cost in standard cell based design flows and how synchoricity will alleviate this problem.

Consider the design space diagram shown in Figure 3 for three different VLSI design styles. These diagrams show six levels of abstraction. The number of solutions exponentially increases with refinement from each level of abstraction to the next. The refinement process starts from the higher abstraction levels (top) and progressively moves to the lower levels (bottom).

In the full-custom design era, the functionality was refined *manually* from the system down to the layout/physical level. Manual refinement does not provide a correct-by-construction guarantee and requires expensive functional verification **(FV)**. Since the final cost metrics are known only after physical design, constraints verification **(CV)** must also be done. The manual refinement, functional, and constraints verification restricted designs to O(10K) gates.

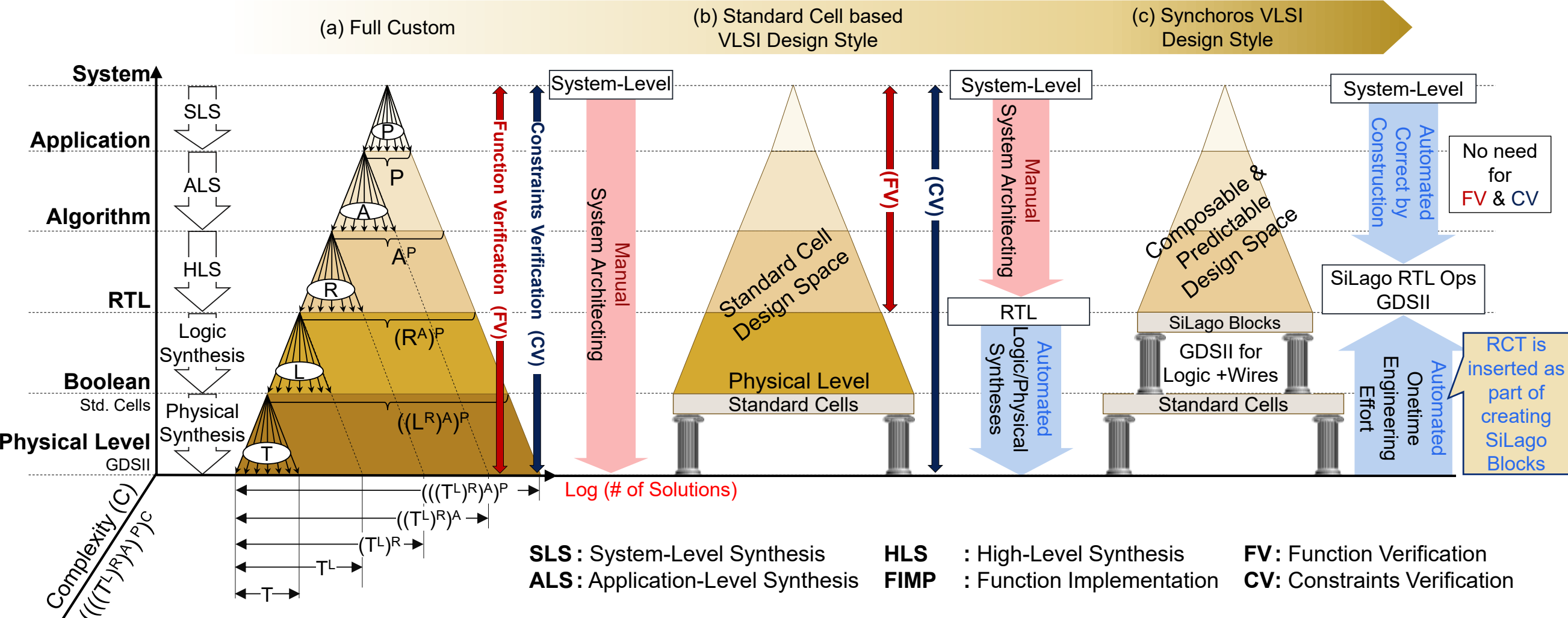

Figure 3. The evolution of VLSI Design Space. Synchoros SiLago blocks exponentially reduce the VLSI design space and make it predictable. This empowers the higher abstraction (algorithm, application, and system) synthesis tools.

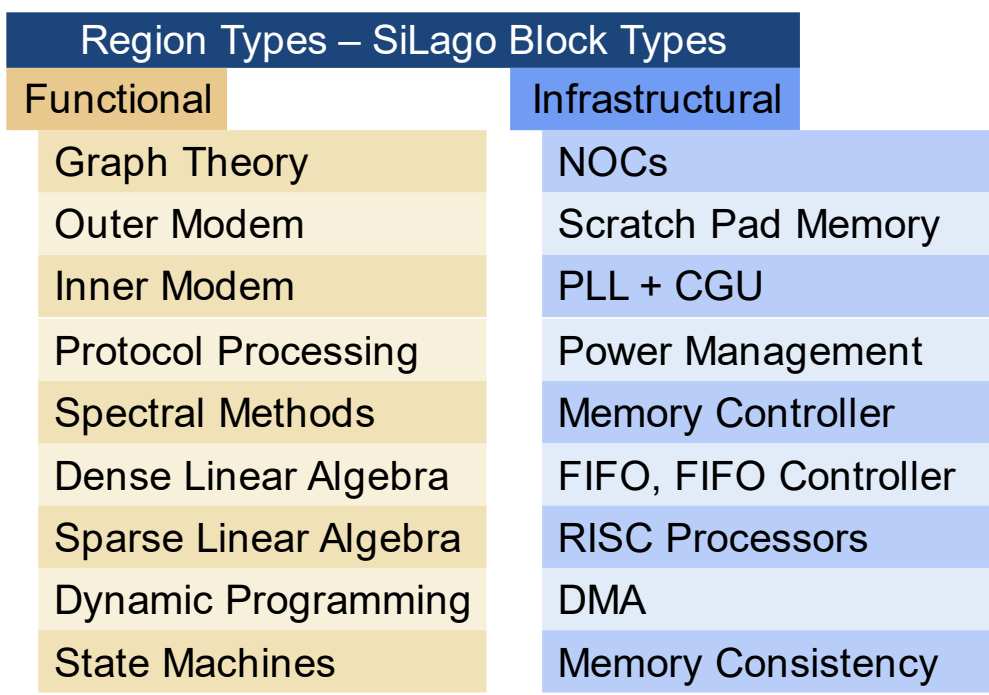

Figure 4. Functional and infrastructural region types.

To go beyond this complexity, the community standardised the VLSI design of standard cells as a one-time effort. By standardising their pitch, the physical design was simplified as well. The standard cells empowered the automation of logic and physical synthesis and the design process from RTL to GDSII. However, the design from system down to RTL has largely remained manual. High-level synthesis, inspired by decades of research, has not become mainstream and has not made much impact [4].

Standard cells made the VLSI design of Boolean level functionality a one-time effort. Still, the placement of the standard cells and the routing of the wires are repeated for every design iteration. Not knowing the impact of wires injects unpredictability that makes even logic synthesis an error-prone process.

The attempt to automate at even higher abstractions has not been successful because the design space is very large, and the physical design and wires are unknown. This unpredictability results in crude estimates that mislead the design space exploration. As a result, system/application/ algorithm level models must be manually refined down to the RTL. Such manual refinement requires functional verification. Additionally, constraints verification is only possible after the physical design has happened. These factors have made the design of custom SOCs or ASICs very expensive and beyond small actors and non-specialists.

The synchoros VLSI design introduces two fundamental innovations. First, it raises the granularity of the atomic physical design objects to micro-architecture level (SiLago blocks) instead of Boolean level (standard cells). Second, by absorbing all metal layer details as part of the SiLago blocks, it obviates the need to synthesise ad-hoc wires; this applies to all wires, functional and infrastructural. This paper focuses on absorbing one such category of wire called the regional clock tree (RCT). Compared to the standard cell-based VLSI design space, the net effect of these two innovations is that the synchoros VLSI design space is a) exponentially reduced and b) it is composable and predictable with post layout accuracy, see Figure 3c. We argue that this makes the synchoros VLSI design space tractable and empowers automation from higher abstractions: algorithm, application (hierarchy of algorithms) and systems (concurrent and communicating applications). The automation is correct by construction and eliminates the need for functional verification. The composition by abutment removes the need to synthesise wires and provides agile and accurate post-layout accurate prediction, eliminating the need for constraints verification.

The synchoros VLSI design style has the potential to eliminate the need for logic and physical synthesis for the end-user. This elimination is analogous to eliminating the SPICE-based simulations when the standard cells were introduced as a replacement for the full-custom design.

The SiLago blocks are heavily customised for specific application domains. There are two broad categories: functional and infrastructural, as shown in Figure 4. The types of SiLago blocks roughly correspond to the Berkeley dwarfs as presented in the Berkley Report on the Landscape of Parallel Computing [5]. With such a library of SiLago block types, it is possible to map any mix of applications as a synchoros VLSI design.

## III. RCT BY ABUTMENT IN SYNCHOROS VLSI DESIGN

This section presents the requirements imposed by synchoricity on the RCT and how these requirements are fulfilled, including creating a valid and predictable design that factors in the process, voltage and temperature **(PVT)** variations and On-Chip variations **(OCV)**.

### *A. RCT requirements in SiLago Design Flow*

The RCT has two types of requirements. One is common to all clock tree syntheses, i.e., to minimise the clock skew and maintain sufficient drive strength to ensure that the slew rate is not violated. The second set of requirements stems from synchoricity; it imposes two additional constraints: predictability and validity. An RCT composed of its fragments should be predictable. The predictability requires that the RCT can only be composed from a finite set of pre-characterised fragments absorbed within the SiLago blocks. We discuss such a timing model in section III.C. The parameters of this model factor in PVT and OCV. The RCT generated by abutment should also be valid, i.e., fulfil the technology design rules and ensure manufacturability.

### *B. RCT Components in each SiLago Block*

In this subsection, we elaborate on the RCT components absorbed within each SiLago block. These components enable abutment, maintain the slew rate and minimise the skew.

*1. Standardised Entry and Exit Points*: RCT fragment in every SiLago block type has a standard entry ($H_{in}$, $V_{in}$) and exit ($H_{out}$, $V_{out}$) points; see Figure 5. Standard implies a fixed location on the specific edge of SiLago blocks and metal layers. The entry and exit points of the RCT fragments in neighbouring SiLago blocks abut to create a valid RCT. Since the neighbours can be in the horizontal, vertical, or both dimensions, the SiLago blocks need entry and exit points on both edges. The RCT can be distributed the top-down and left-right, or bottom-up and right-left. The choice of orientation depends on the corner at which the global clock tree **(GCT)** enters. The GCT entry point depends on the floor planning of the global NoCs during the synthesis from higher abstractions.

*2. Multiplexed and buffered horizontal and vertical chords*: These components select the RCT input and output and maintain the slew rate. Selecting the input implies selecting the $H_{in}$ or $V_{in}$, as shown in Figure 5. The two inputs are fed to an OR gate. Only one of the inputs can be a clock, and the other is set to zero when configuring the RCT. Selecting the output implies selecting if the RCT is propagated to the right exit (Hout), the bottom exit ($V_{out}$), or both. The unselected exit is grounded using two AND gates. These gates also serve as the drivers to maintain the clock's slew. Depending on the two AND gates' configuration, the variants of chord delay, TRCT_chord, is selected; see Figure 5.

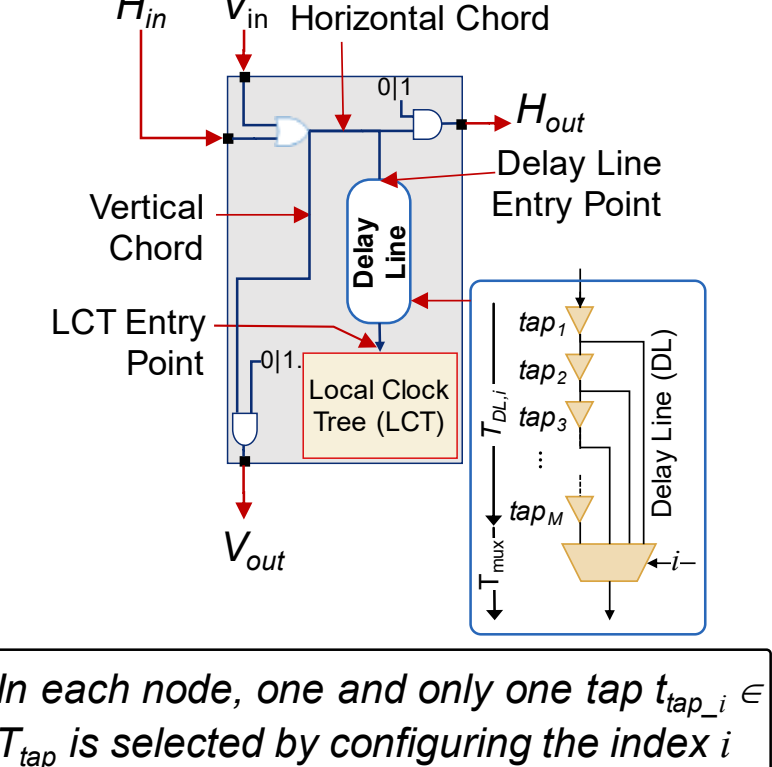

| SiLago Block Delay Notation | |
|---|---|
| $T_{RCT_chord}$ | $T_{V_to_V}$, $T_{V_to_H}$, $T_{H_to_H}$, $T_{H_to_V}$<br>Four variants of delay through horizontal and vertical chords |
| $T_{V_to_DL}$, $T_{H_to_DL}$ | Delay of the wire connecting the $V_{in}$ or $H_{in}$ and the Delay Line Entry Point |
| $T_{DL_i}$ | Delay imposed by delay line depending on the index $i$ |
| $T_{mux_i}$ | Propagation delay through the mux depending on the $i$ index |
| $t_{tap_i}$ | Delay between the $V_{in}$ or $H_{in}$ and the LCT Entry Point associated with the selected tap $i$ of the Delay Line<br>$t_{tap_i} = [T_{H_to_DL} \mid T_{V_to_DL}] + T_{DL_i} + T_{mux_i}$ |
| $T_{tap}$ | Set of delays associated with taps in the Delay Line<br>$\{t_{tap_0}, \ldots, t_{tap_i}, \ldots, t_{tap_(M-1)}\}$ |

Figure 5. Components of RCT fragment and their notations.

*3. Programmable Delay Line* ***(PDL):*** PDL adjusts the delay to the local clock tree **(LCT)** entry point; see Figure 5. The delay is adjusted according to the SiLago block's position in a region instance with respect to where the GCT enters the region. The objective of adjusting the delay is to minimise the skew of the clock's arrival at the LCT entry points. The delay is adjusted by selecting a tap, with index *i,* in the delay line. The selected tap $i$ introduces a delay $t_{tap_i}$ between the SiLago Block's RCT entry point, i.e., $H_{in}$ or $V_{in}$, and the LCT entry point.

The RCT fragments' design, in terms of the components described above, is identical for all SiLago block types. However, the dimension of the three components depends on the type of the SiLago block. The entry and exit points of the SiLago blocks are like the Lego studs. The position of these points is at a standard offset from top-left and bottom-right corners, irrespective of the type of SiLago blocks. This standardisation makes it possible to abut SiLago blocks of different types. By having standardised offsets from the corners, the spatial composition of blocks becomes feasible.

The length of the horizontal and vertical chords is adjusted to match specific SiLago block types' dimensions. The drive strength of the three gates used for configuring the propagation path is also dimensioned to fit their respective loads. These loads are not arbitrary but known a priori. This is possible because a SiLago block can have only a finite number and types of neighbours. When a SiLago block type is designed, its connection to all possible neighbour block types is characterised to ensure that the electrical connection among them is valid.

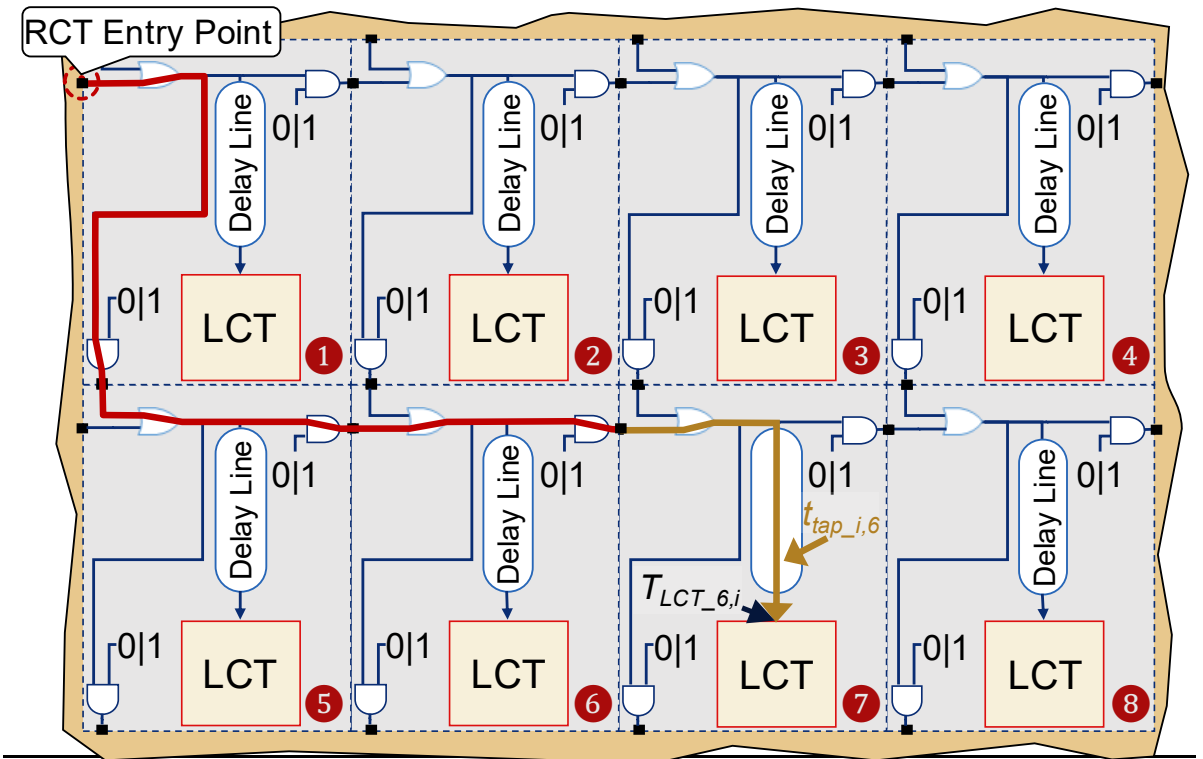

Figure 6. RCT delay model components in an example region instance.

## *C. Regional Clock Tree Delay Model*

The RCT delay model enables post-layout accurate predictability of electrical and temporal properties of an arbitrary RCT created by the abutment of its fragments. This model is used to decide how large a region instance can be and select the taps in the delay line to minimise the skew among the LCT entry point (Figure 5) of the SiLago blocks.

The delay model captures the RCT latency from the entry point of a region instance to the LCT entry point of the SiLago blocks in a region. Figure 6 presents an example of how the delay model is applied. It illustrates a region instance with 8 SiLago blocks, where the RCT entry point is on the top left corner. Each block will have a unique RCT arrival time, notated as $T_{LCT_x,i}$. The subscript $x$ identifies the block id, and the subscript $i$ identifies the programmable delay line's selected tap index. The programmable line aims to make $T_{LCT_x,i}$ as equal as possible for every block $x$. The $T_{LCT_x,i}$ has two components, as shown in Eq. (2).

a. The first component is the propagation delay in the intervening RCT chords between the RCT entry point in a region instance and the entry to a SiLago block $x$. This delay is denoted by $T_{prop_x}$, where $x$ is the SiLago block id, Eq. (1). In Figure 6, $x=7$ and $T_{prop,x}$ is shown as the thick red line, colouring the chords in blocks 1, 5, and 6.

b. The second component is the delay imposed on the RCT by the delay line in the SiLago block $x$. This delay is represented by $t_{tap_i,x}$, where $i$ is the tap index. In Figure 6, this delay component is shown as $t_{tap_i,7}$ in a thick golden line in block 7. The $t_{tap_i,x}$ is further divided into three sub-components, as shown in Figure 5.

$$T_{prop_x} = \sum_{nodes\ previous\ to\ x} T_{RCT_chord} \tag{1}$$

$$T_{LCT_x,i} = T_{prop_x} + t_{tap_i,x} \tag{2}$$

The SiLago design flow involves two phases. The blocks are designed, verified, and characterised as a one-time effort in the first phase. A partial list of outputs that are produced in this phase are:

1. The RCT fragments are analysed and characterised to extract the parameters used in the RCT timing model presented above.
2. The fragments of the regional NoCs are also analysed and characterised in the same manner as the RCT's. Examples of such fragments are shown in Figure 7, denoted as $a_r$ and $b_r$. The timing analysis for such inter-SiLago flop-to-flop timing paths is done for each SiLago block type in conjunction with all possible valid neighbour block types. Such timing analysis is used to characterise the delay of the wire fragments. The validation of such timing paths cannot be done until an RCT is created when a design is composed in terms of the SiLago blocks during the second phase involving the end-user. The inter-SiLago timing path example shown in Figure 7 spans to the nearest neighbours. In practice, such timing paths can exist in a small continuous cluster of SiLago blocks in a region instance. The size of the cluster depends on the NoC used to connect the SiLago blocks. The SiLago blocks do not communicate over ad hoc wires, the way standard cells do, but over a structured NoC based interconnect [6].
3. The intra-SiLago, i.e., local timing paths (flop-to-flop), are analysed by the static timing analysis **(STA)** that is part of the standard-cell based design flow. Examples of these paths are noted as $a_l$, $b_l$, and $c_l$ in Figure 7; the subscript *l* stands for local.

In the second phase, a library of characterised and abutment ready SiLago blocks is used to create more complex designs for algorithms, applications, and systems. The higher abstraction syntheses tools automate the synthesis process to create such complex designs [7]. In this phase, the inter-SiLago block timing paths are analysed. The RCT timing model plays a central role in this higher abstraction timing analysis by predicting the skew between the RCT entry points into each SiLago block in a region instance. This skew is the only unknown when composing a design in terms of the SiLago blocks. A method presented in the next section IV is used to minimise the skew. Knowing a) the clock period, b) the delay over the inter-SiLago regional NoC timing paths from its pre-characterised fragments like $a_r$ and $b_r$ and delays in the flops, and c) the skew predicted by the RCT timing model, it is possible to validate the inter-SiLago timing paths as timing clean. The RCT model can predict the skew as accurately as a static timing analysis **(STA)** tool. We validate this claim in section V.C. The RCT timing model is never used explicitly by the end-user. However, it is used by the higher abstraction synthesis tools to a) decide the taps in the delay lines, b) validate the inter-SiLago timing paths, and c) calculate the largest feasible size of region instance. The RCT timing components' characterisation and the inter-SiLago timing paths are like the properties of standard cells and wires in a specific technology. The end-user does not explicitly use these properties, but they empower logic and physical synthesis tools. Likewise, the SiLago block timing models empower synthesis from abstractions higher than RTL.

### D. *Maximum feasible size of region-instances*

The RCT model introduced in section III.C is also used to decide a region instance's maximum size. The number of taps and their ability to compensate for the monotonous increase in $T_{nat_x}$ dictates the maximum size of region instances allowed. This is formalised in Eq. (3). The value of $x$ that fulfils this inequality decides the furthest node, $N$, from the RCT entry point and thereby the maximum feasible dimension of the region instance. Note that the maximum size does not need to be an optimal choice during higher abstraction synthesis. However, these synthesis tools need to know the upper bound.

$$max(t_{tap_i}) - min(t_{tap_i}) \leq max(T_{nat_x}) \qquad (3)$$

### E. *On-Chip Variations*

On-chip variation **(OCV)** is a well-known challenge that impacts timing. There are three main approaches in use today for the standard-cell based design flows [8], [9]: a) single deration per process, voltage, and temperature **(PVT)** point, b) advanced OCV, and c) parametric OCV. All these methods rely on standard cell libraries to define how much variation can be expected during manufacturing. The EDA tools take the variations into account during the STA to ensure that each timing path is designed with sufficient margins so that no violations can occur in any process or operating corner.

The synchoros VLSI methodology implicitly factors in the OCV. The incorporation happens when the SiLago blocks are designed, analysed, and *characterised* using the OCV aware standard-cell based design flows. The SiLago blocks are characterised at all delay corners for two reasons. The first reason is to ensure that all its local intra-SiLago timing paths are timing clean in these corners. The second is to extract timing parameters for the fragments of the wires that abut to create inter-SiLago wires when a SiLago based design is composed. These timing parameters are bounded by given OCV ranges to ensure their validity in the various corners. As a result, the OCV ruggedised parameters are used for the RCT timing models and the inter-SiLago NoC fragments when doing the inter-SiLago timing path analysis. This ensures that all the timing paths are analysed to the same standard as the conventional EDA flows, though at a higher abstraction.

## IV. Optimal Tap Selection in Delay Lines

The optimisation problem of tap selection is formulated as the assignment of the tap index in each node x, which would minimise the absolute difference among $T_{LCT_x,i}$. We first develop a straightforward, locally optimum cost function and later a globally optimum cost function.

To formally define these cost functions, we introduce some notations and conventions. Node IDs are *1…N,* and the delay line tap indices are *1…M*. The first tap, $i=1$, imposes the minimum delay and the last tap, $i=M$, the maximum delay. In other words, $t_{tap_1}=min(t_{tap_i})$ and $t_{tap_M}=max(t_{tap_i})$. We remind here that all nodes have the same delay line with the same number of taps and delays associated with each tap. The *ID* of the node where RCT enters a region instance is by convention *1,* and it has $T_{nat_1}=0$ since there are no previous nodes. The node *ID N* is reserved for the furthest node, i.e., $T_{nat_N}= max(T_{nat_x})$. The tap index in node *N* is fixed to $t_{tap_1}$ to have minimal latency, $T_{LCT_N,1}$. This selection is made for two main reasons. The first is to minimise the insertion delay to the blocks. The second is to minimise the number of buffers used in each delay line, reducing the power consumption of the clock tree. As a result, the search space excludes the tap index space in node $N$, and it is fixed to 1.

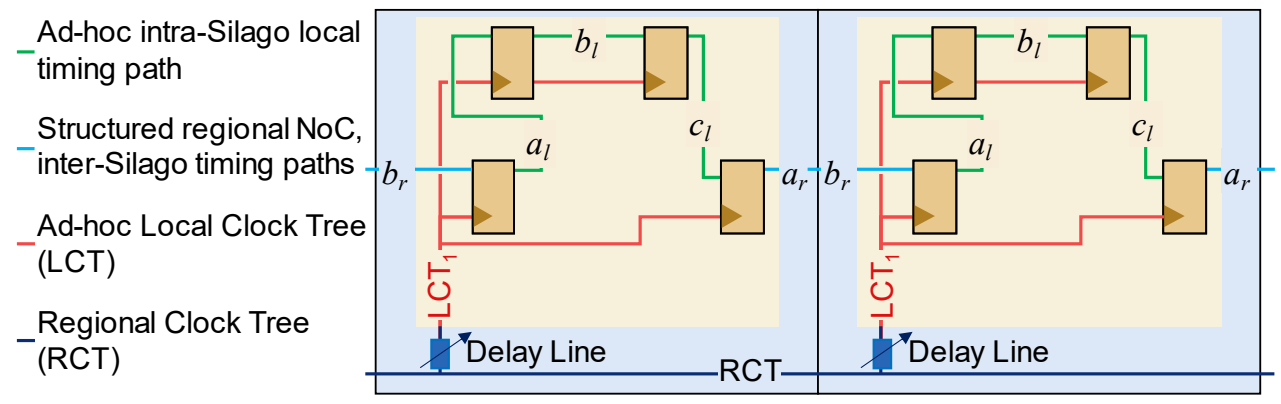

Figure 7. Local & regional clocks and timing paths.

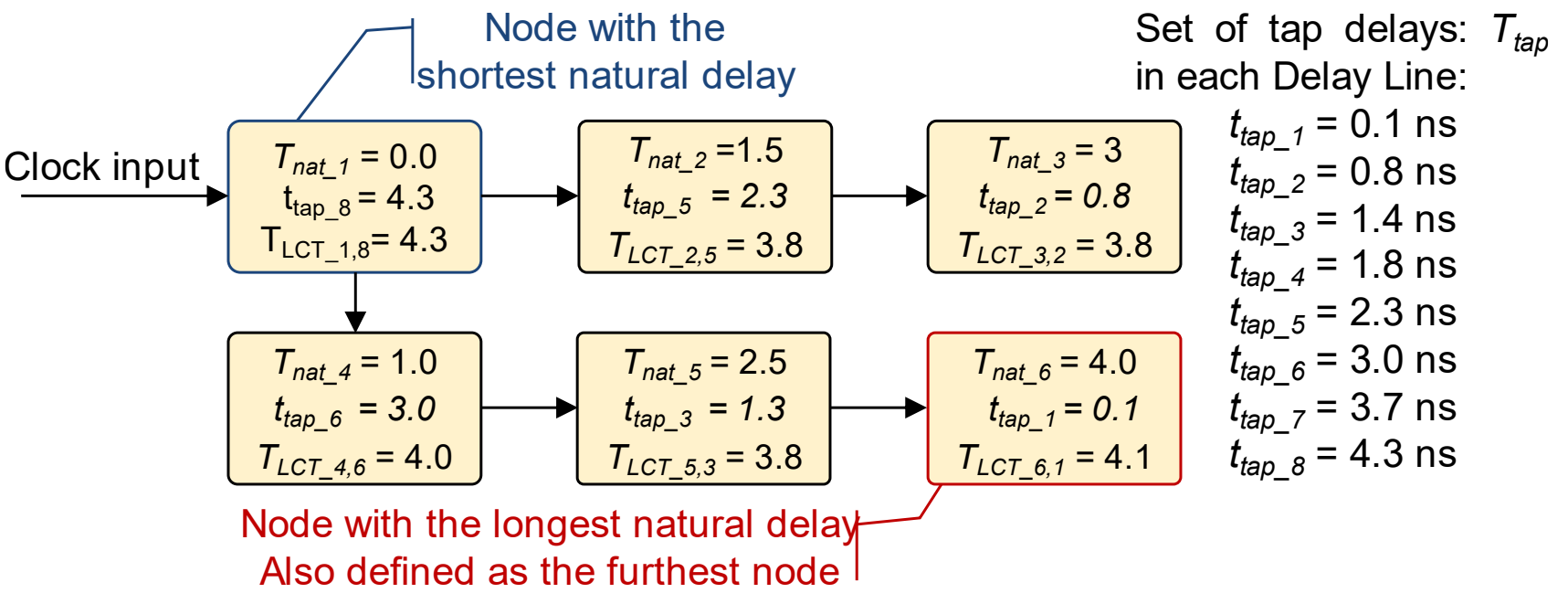

Figure 8. A motivational example that illustrates the tap index selection for the locally optimum solution.

### A. Locally Optimal Solution

The locally optimum cost function in Eq. (4), $L$ quantifies the mean of the absolute differences between $T_{LCT_x,i}$ and $T_{LCT_N,1}$, where $x=1...N-1$.

Since $L$ is a sum of absolutes, the minimality of $L$ can only be guaranteed if each term in the summation is also minimal. This minimality, in turn, can be guaranteed by visiting each of the $1...N-1$ nodes and sweeping through the $M$ taps to find the index that gives the minimal absolute difference with respect to the reference node $N$. This recipe for finding the locally minimal solution $I_{LM}$ is formalised in Eq. (5).

$$L = \frac{1}{N-1} \sum_{x=1}^{x=N-1} \left| T_{LCT_x,i} - T_{LCT_N,1} \right| \quad (4)$$

$$\begin{aligned} I = \{ i \mid 1 \le i \le M \quad & AND \\ 1 \le x \le N-1 \quad & AND \\ T_{LCT_x,i} - T_{LCT_N,1} = \min_{1 \le j \le M, j \ne i} & \left( T_{LCT_{x,j}} - T_{LCT_{N,1}} \right) \} \end{aligned} \quad (5)$$

To conclude, $L$ will be minimal if we replace $T_{LCT_x,i}$ with $T_{LCT_x,k}$ where $k=I_{LM}(x)$. The complexity of a locally optimum solution is $O(N \cdot M)$. This is evident from Eq. (4), where $N$-$1$ nodes are evaluated, and there are $M$ taps to be evaluated in each node.

What makes $L$ local is that it minimises the mean with respect to a single node – the furthest node. If the reference node is changed to some other arbitrary node $K$, there is no guarantee that $I_{LM}$ would give a minimal $L$. In short, $L$ related to any node would prioritise the needs of a single node, potentially at the expense of other nodes.

Before presenting the global cost function, we would like to compliment the concepts introduced above with a concrete example. This example is also used with the global cost function to highlight the difference.

A SiLago region instance can be modelled as a DAG (*Directed Acyclic Graph*) laid out as a two-dimensional mesh shown in Figure 8; this example is a subset of the example shown in Figure 6 with 6 instead of 8 nodes. Each node in the DAG represents a SiLago block, and the edge to the nearest neighbour represents $T_{RCT_chord}$. Without loss of generality, let us simplify the four variants of $T_{RCT_chord}$ to two: horizontal and vertical, with 1.5 and 1.0 units of delay, respectively. Each node $x$ is annotated with its node-ID and the values of delays introduced in 1 and 2. The values in the nodes reflect the assumption that the RCT enters the region instance at the top-left corner and propagates in a top-down, left-right fashion. Each node is equipped with a delay line with $M=8$ taps. The delay associated with each tap is shown on the right side in Figure 8. In this example, the taps are selected to have a minimal $L$ corresponding to the furthest node 6.

The Eq. (4) is used to calculate the $L$ in the example given in Figure 8 and is equal to:

$$L = {}^{1}/_{5} \sum_{x=1}^{x=5} \left| T_{LCT_x,i} - T_{LCT_6,1} \right| =$$
$${}^{1}/_{5}\,(0.2 + 0.3 + 0.3 + 0.1 + 0.3) = {}^{1.2}/_{5} = 0.24$$

### B. Globally Optimum Solution

A global cost function would take into account the needs of all nodes. The $L$ in Eq. (6) is such a cost function and quantifies the sum of differences of $N$-$1$ nodes with respect to each of the $N$ nodes as the reference; $y$ replaces $N$ in 3 and is swept over $1...N$.

$$L = \frac{1}{N(N-1)} \sum_{y=1}^{y=N} \sum_{x=1}^{x=N-1, x \ne y} \left| T_{LCT_i,x} - T_{LCT_i,y} \right| \quad (6)$$

The global nature of Eq. (6) implies that it is no longer sufficient to select the tap index in each node *independently* of the tap index selection in other nodes. Since there are $N$ nodes and each node has $M$ taps, there are $M^N$ possible configurations of the delay lines, $I_{conf}$, in a region instance:

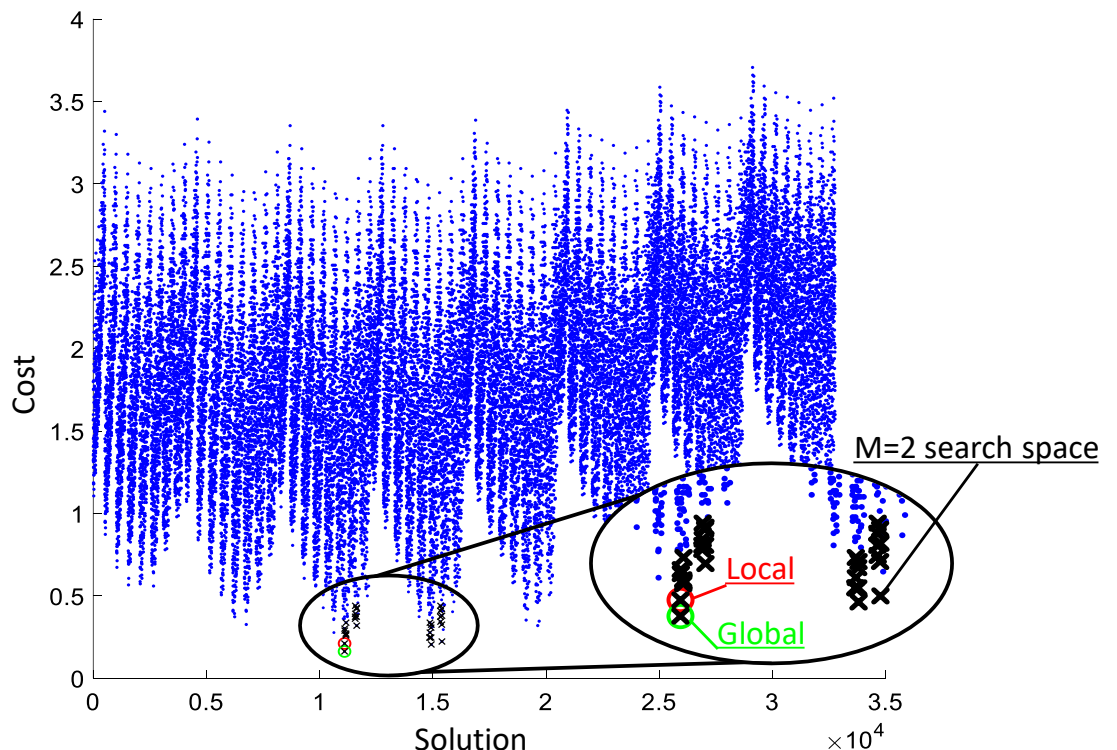

Figure 9. Solution points in globally optimum search space with a red dot highlighting the globally minimum solution.

$$I_{conf} = \{I_1, I_2 \dots I_k \dots I_{M^N}\} \text{ where}$$
$$I_k = (i_1, i_2 \dots i_x \dots i_N) \text{ and } i_x \in \{1, 2 \dots M\}$$

Note that $I_K$ is a sequence and not a set; the elements of this sequence i) have an order $1 \dots N$ and ii) can have duplicates. Next, we define $L(I_k)$ for the candidate solution $I_k$:

$$L(I_k) = \frac{1}{N(N-1)} \sum_{y=1}^{y=N} \sum_{x=1}^{x=N-1, x \neq y} \left| T_{LCT_ix,x} - T_{LCT_iy,y} \right|$$
$$where\ ix = I_k(x)\ and\ iy = I_k(y)$$

The globally minimal solution $I_{GM}$ is then defined as follows:

$$L(I_{GM}) = \min_{I_k \in I_{conf}} L(I_k)$$

We next apply the above recipe to find the globally optimal solution for the same problem used in the previous section. We have plotted all $8^5$ different solutions in the globally optimum search space, Figure 9. The green circle identifies the globally minimal solution: $I=\{7, 5, 2, 6, 3, 1\}$. The cost of the globally minimal solution using Eq. (6) is $0.173$. If we instead take the locally optimal solution presented in the previous subsection and apply Eq. (6), we get the result of 0.24. This result proves the benefit of adopting the globally optimum cost function, formalized in Eq. (6).

The globally optimal solution has an exponential complexity of $O(N^2 \cdot M^N)$. The $N^2$ factor represents the complexity of calculating the cost function. In the experimental platform, we report in section V, $M=32$. Even for a small region instance of just ten nodes, the complexity is of the order of ~$10^{17}$ evaluations of absolute differences. Fortunately, this design space can be easily pruned to a scalable size without sacrificing global optimality. We next justify how $M$ can be pruned down to $2$, i.e. $M=2$ independent of the dimension of the delay line. We also motivate why we can replace $N$ with $N'$, where $N' \ll N$ and $N'$ is independent of the size of the region instance.

### C. Pruning the Search Space of the delay-line tap selection

This subsection justifies the basis for dramatically pruning the configuration space.

#### 1) Justifying M=2

Ideally, we would like $L$ to be zero. Since these cost functions are the mean of absolute differences, each term should be reduced to zero for them to be zero. The two $T_{LCT}$ terms in absolute difference expression must be equal for this condition to be true. Since the $T_{nat}$ component in $T_{LCT}$ is invariant, the $t_{tap}$ should have an exact delay to make the two $T_{LCT}$ equal. This delay is called the ideal $t_{tap}$ delay. We formalise this ideal $t_{tap}$ delay in Eq. (7).

$$t_{tap_ideal,x} = T_{LCT_N,1} - T_{nat_x} \qquad (7)$$

Such an ideal tap delay is not possible in practice because it requires an infinitely divisible delay line. However, the ideal tap delay helps identify the two contiguous tap indices whose delays bracket the ideal tap delay. This is illustrated in Figure 10, and as can be seen, $t_{tap_ideal}$ is the basis for finding the two taps $i$ and $i+1$ whose delay brackets the $t_{tap_ideal}$ in node $x$. We can repeat this process for all $N-1$ nodes. As a result, each node's lower and upper bound taps i and i+1 will be identified.

In conclusion, the complexity reduces to $O(2 \cdot N)$ and $O(N^2 \cdot 2^N)$ for locally and globally optimum solutions, respectively. $2^N$ is still a large factor, and we next justify why this can be pruned down by replacing N with N'≪N.

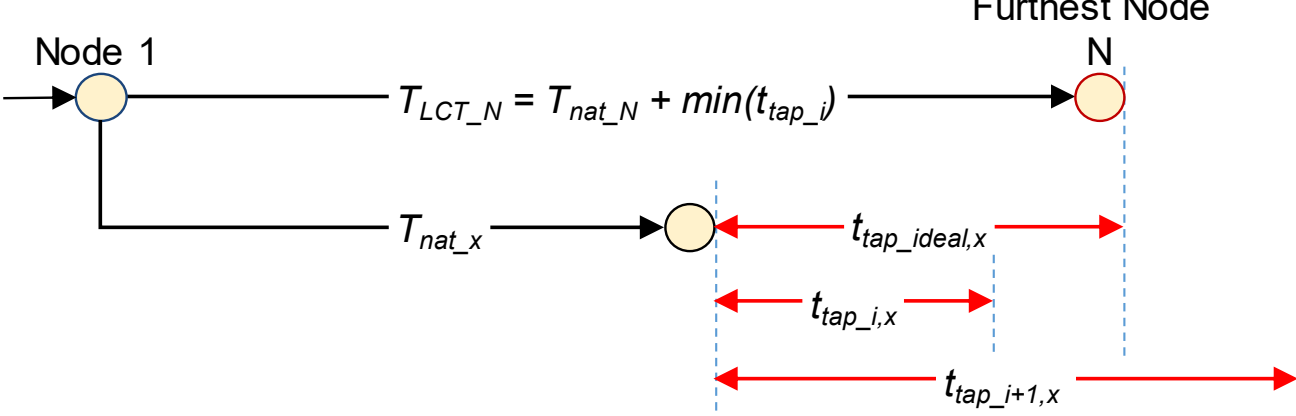

Figure 10. $t_{tap_ideal}$ helps find the two neighbouring indices in each node that needs to be considered during optimisation

#### 2) Justifying $N' \ll N$

A combinational path implies a connection from a register's output to a register's input composed of wires and combinational logic. The clock skew between such register to register paths should be minimised. The cost functions specified in Eq. (4) and Eq. (6) attempt to minimise such skews, i.e., the absolute difference in $T_{LCT}$ between the furthest node pairs (like $1$ and $N$) and between the nearest neighbours. Such optimisation would be necessary if a combinational path connects every node pair in a region instance. This is not the case for SiLago region instances composed using the regional NoCs and is the basis for replacing $N$ by $N' \ll N$.

To understand this, let us revisit the local NOCs: the intra-region-instance, structured interconnect scheme in synchoros VLSI region types, see section III. These local NOCs serve two purposes. The first is conventional; to allow SiLago blocks in a region instance to communicate with each other. The second embodies the spirit of synchoricity and allows the functionality hosted by individual SiLago blocks to be clustered to provide variations in function, capacity and degree of parallelism. Standard-cell-based VLSI design achieves this objective by synthesising ad-hoc wires to cluster the standard cells. In synchoros VLSI design, these wires pre-exist as fragments of local NoCs in SiLago blocks. The region-wide local NoCs emerge as a result of the abutment. However, in synchoros VLSI design, the combinational paths between SiLago blocks are restricted to a small window. The window slides by a certain stride to cover the entire region instance, as shown in Figure 11. This structure implies that two SiLago blocks located in different windows do not have any combinational path between them. As such, there is no need to worry about clock skew between them. Communication between such SiLago blocks is possible but involves multiple hops/cycles, i.e., the path is pipelined.

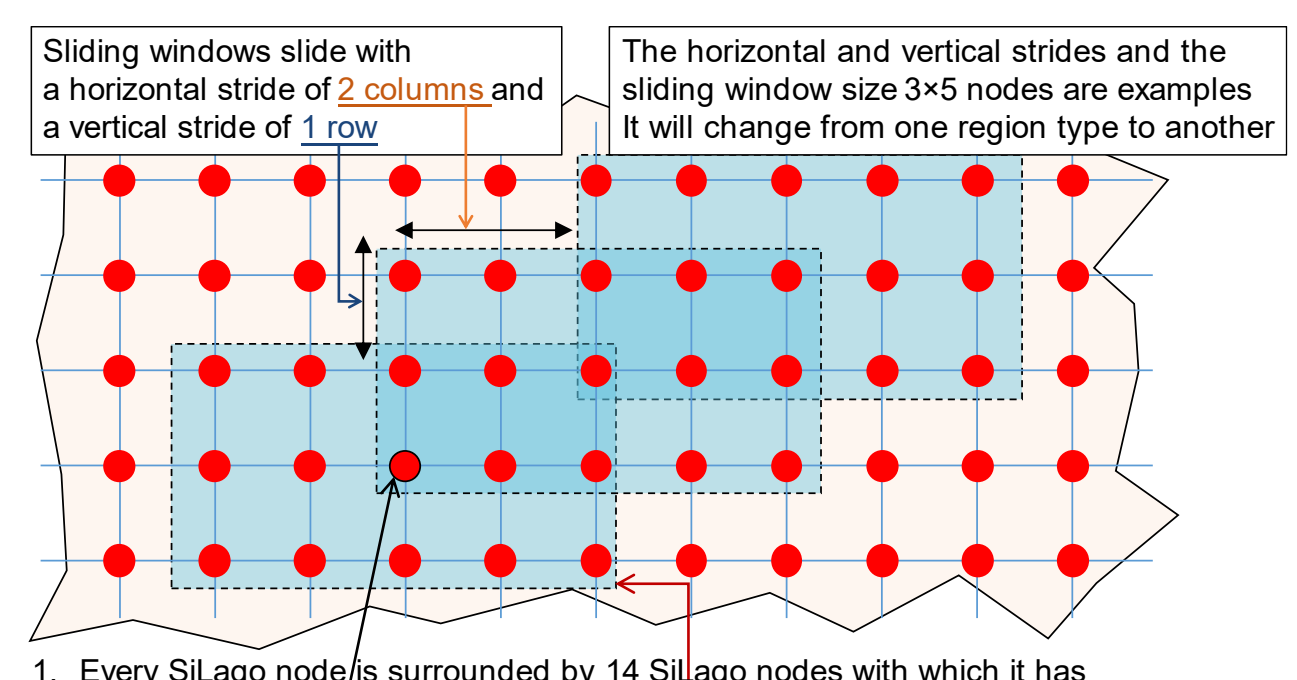

Figure 11. Every node has combinational paths to N'-1other nodes in the span of the sliding window creating a need to minimise $T_{LCT}$ among nodes with such sliding windows rather than the whole region instance.

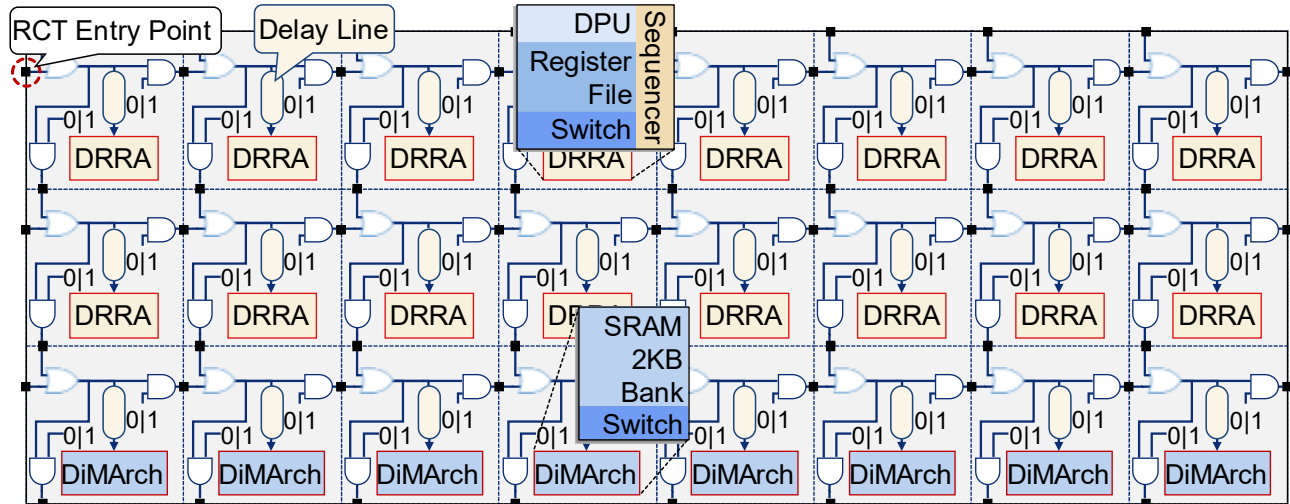

Figure 12. The experimental region instance.

Since every node is not connected to every other node but only to nodes surrounding it in a sliding window, the problem can be simplified. The absolute difference in arrival time that RCT needs to minimise is between every node $x$ and a small set of nodes surrounding it in a sliding window. The nodes inside the sliding window are reachable from $x$ over a combinational path. In the example shown in Figure 11, this small set of nodes $N'$=15. In the synchoros VLSI region type called DRRA, $N'$=14, see [6]. The sliding window size decides N' and is independent of the size of the region instance.

## V. Experiments and Results

This section presents the experimental results that validate the claim that the RCT generated by abutment is: a) valid, i.e., it is guaranteed to be timing and design rule check **(DRC)** clean, and b) predictable, i.e., properties of the generated RCT are known with post layout accuracy without having to do STA at the physical level, as it is done in the standard cells design flow.

Three experiments were performed to validate these claims. The first experiment reports the RCT delay model results and its properties, as discussed in section III.C. The second experiment uses the RCT delay model to predict the properties of the RCT. The predicted values are validated against the values analysed by commercial EDA tools. This experiment's side effect is that the RCT created by abutment is validated as timing and DRC clean by the commercial EDA tools. We also demonstrate that the predictability is scalable. The third experiment benchmarks the RCT properties generated by abutment against a functionally equivalent RCT generated by the EDA tools. The results prove that the synchoricity and abutment have negligible overheads in terms of typical cost metrics for clock tree: area, wire length, switching capacitance, skew, and average trunk slew.

### A. Experimental setup

All experiments have been done in a 40nm technology node, and the results have been validated using commercial EDA tools. These tools have been used to a) implement synchoros SiLago blocks, including the RCT fragments, b) validate the claims that the RCT generated by abutment is predictable, and timing and DRC clean, and c) demonstrate that the benefits of synchoricity and abutment do not degrade the quality of RCT. EDA tools for points *b* and *c* above are solely used for research purposes to validate the claims and are not part of the synchoros design flow intended for the end-user.

The proposed scheme for generating the RCT by abutment and the state-of-the-art hierarchical EDA flow is applied to the same experimental design. The design is a composite region instance of two different types of SiLago blocks. One is called Dynamically Reconfigurable Resource Array **(DRRA)** [6], [10], [11], and a second fabric is called Distributed Memory Architecture **(DiMArch)** [12]–[14], Figure 12.

The region instance has 24 SiLago blocks corresponding to ~1.5 million NAND gates and 16 kBs of SRAM or 4 mm$^2$ in a 40 nm technology node. The design's size is compatible with the expected typical size of synchronous region instance, for which the RCT is generated by abutment. To establish the method's scalability, we also test a ~4 million gate region instance design and show that the predictability is unaffected.

### B. RCT Model

An RCT fragment with 32 taps in the delay line was incorporated into the DRRA and DiMArch SiLago blocks. These blocks were hardened to be synchoros and abutment ready. The RCT model parameters were extracted using static timing analysis **(STA)** from the post layout data and factoring in PVT and OCV. We emphasise that this STA is a one-time engineering effort to characterise SiLago blocks and is not seen by the end-user. The SiLago blocks on the edges have slightly different values of $T_{RCT_chord}$ compared to the ones in the middle. These differences exist because there are minor differences in the interconnect and their layout. The 32 taps in the delay line can insert delays from 1.07 ns to 6.2 ns, with a step size of 0.16 ns. The maximum size of the region that can be supported is nine columns. A delay line with more taps or a larger step size can support larger regions. The RCT timing model is based on the extraction of timing using EDA tools that factor in PVT variations.

### C. Predictability, Validity, and Scalability

The RCT created by abutment is predictable and valid. The property that we predict is the arrival times of RCT at each LCT entry point, the $T_{LCT\,x,i}$. We measure this property using two methods, see Figure 13. The first method uses the RCT delay model and the higher abstraction timing analysis to analyse the inter-SiLago timing paths composed of the pre-designed and pre-characterised wire fragments absorbed in the SiLago blocks. This higher abstraction timing analysis for RCT is expressed as in Eq. (2). The second method uses EDA tools to measure the same delay. The results of these two methods are compared to establish the accuracy of the RCT model's prediction with EDA tools-based measurements as the benchmark. The second method is only required to validate the results of the first and serve as a benchmark. It is not part of the SiLago design flow. This experiment's side-effect is that the RCT created by abutment gets certified as being timing and DRC clean by the commercial EDA's analysis tools.

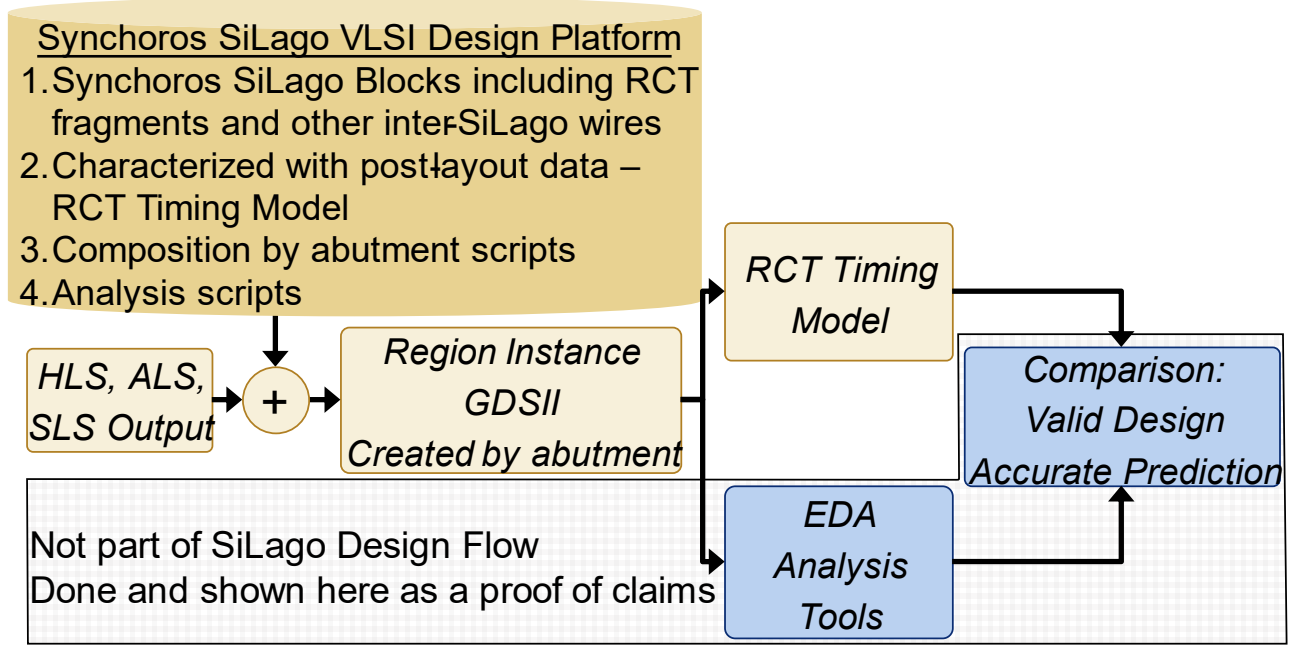

Figure 13. Experimental flow to compare validity and predictability.

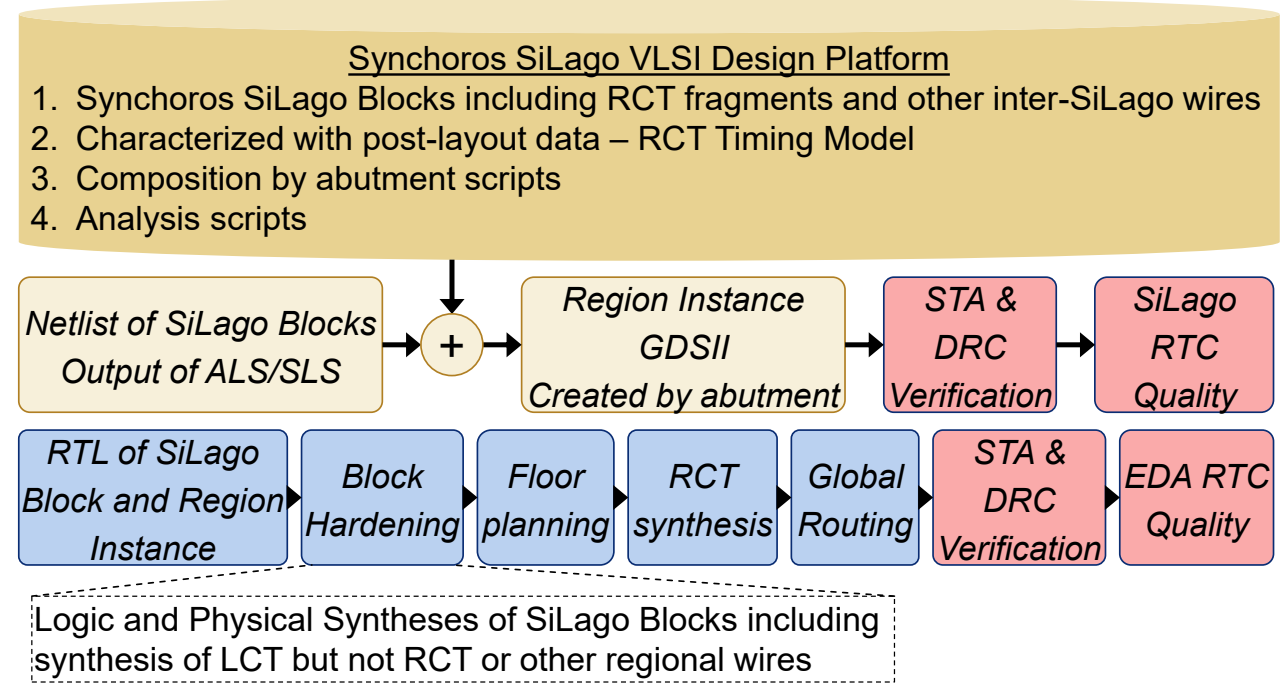

Figure 14. Experimental setup comparing the SiLago RCT with the one generated by the EDA flow.

The worst-case skew, the difference between two LCT entry points, is 129 ps. This difference is small enough to be easily absorbed by the slack margin with which the SiLago blocks are synthesised. We remind here that the SiLago blocks do not communicate over ad-hoc wires synthesised for every design instance and iteration, as is the case with standard-cell based designs. The inter-SiLago communication in a region instance happens over regional NoCs. The RCT delay model's predicted values are almost identical to those analysed by the EDA tools. The worst-case error compared to the EDA tools is 1.5 ps, and its RMS is 0.0005ps. This difference is because our experimental setup does not have an infinite ground plane. This results in different design parts experiencing different coupling with long signals, like the reset and the ground plane. The result proves that the RCT delay model is sufficiently accurate. The higher abstraction synthesis tools do not need to do static timing analysis at the physical design level. To establish the RCT's scalability and predictability with increasing complexity, we experiment with a larger design of 20 columns. The predictability of the larger design is as accurate as that of the smaller design.

### D. *SiLago RCT compared to EDA RCT*

Here we present results that show that the RCT generated by abutment has comparable cost metrics to an ad-hoc RCT generated by EDA tools. To compare the results, we applied two different flows and experiments, see Figure 14. The first uses the Silago flow, and the second is based on commercial EDA tools to generate a functionally equivalent RCT. In the EDA flow, we first harden each DRRA cell as a leaf node. This hardening phase includes the synthesis of the local clock tree. The next step is to floorplan the design in terms of leaf nodes and synthesises the ad-hoc wires to connect and clock these leaf nodes. The clock that is synthesised in this step to drive the local clock trees is functionally equivalent to the RCT generated by abutment. The commercial EDA tools then analyse both designs to compare the properties of the two functionally equivalent RCTs. The results are shown in TABLE I. As can be seen, the values of critical parameters are comparable. The standard cell area refers to standard cells dedicated to the RCT components shown in Figure 5, including MUX/AND/OR gates. The capacitance of SiLago RCT is ~2.5% of the total clock tree capacitance (RCT + Local Clock Tree-**LCT**). The corresponding EDA RCT capacitance is 0.4% of the whole clock tree. The overhead of the SiLago compared to the EDA RCT, as a percentage of the entire clock distribution (RCT + LCT), is 2.1%.

TABLE I.
EDA RCT VS SILAGO RCT

| | SiLago RCT | EDA RCT | % Change |
|---|---|---|---|
| Wire Length [μm] | 422932.1 | 415263.9 | 1.8 % |
| Standard Cell Area [μm²] | 19472.6 | 18276.5 | 6.5 % |
| Average Trunk Slew [ns] | 0.062 | 0.090 | 31.1 % |
| Total Capacitance [pf] | 129.9 | 127.2 | 2.1 % |
| Avg. Diff. in Arrival Time (LCT Entry Point) [ns] | 0.04299 | 0.0337 | 27.6% |

The SiLago RCT achieves a comparable arrival time of RCT at LCT entry points compared to the commercial EDA RCT. The SiLago RCT has a slightly better slew rate at LCT entry. The average and absolute difference in slew at different points in trunks for the entire (RCT+LCT) clock tree is also comparable and within the limits of the technology rules. The above experiment and the reported results establish that RCT generated by abutment has comparable quality to the one generated by the commercial EDA tools. The difference is that the EDA tool generated RCT is ad-hoc and synthesised anew for each design instance and iteration. Notice that the EDA RCT's irregularity in Figure 15 results from the attempt to factor and reuse the buffers. In *contrast*, the synchoros RCT is regular. As discussed in section III, the EDA RCT's ad-hoc nature and its irregularity violate the synchoros design style's requirement on RCT.

The synchoros RCT by abutment is regular, as shown in Figure 15b. It has three main branches corresponding to two DRRA and one DiMArch row. Each branch has eight leaf nodes, and each leaf node has the same RCT structure. The regularity of the SiLago RCT enables abutment. Its regular structure, together with its absorption in the SiLago blocks as a pre-synthesised and characterised structure, enables predictability. The proposed method's main drawback is the clock latency, which is higher than the EDA RCT's, as shown in Figure 15. The increased variation implied by the higher latency is factored in the PVT variations. The higher latency is defended by its potential to improve design productivity by enabling a predictable RCT and a composition by abutment scheme. We note that when we transitioned from full-custom to standard-cell-based designs, we accepted significant overheads to improve design productivity. Likewise, 90% of the timing paths use only 20% of the clock period [2]. But the discretisation of time by clock ticks enables composition by abutment of synchronous designs.

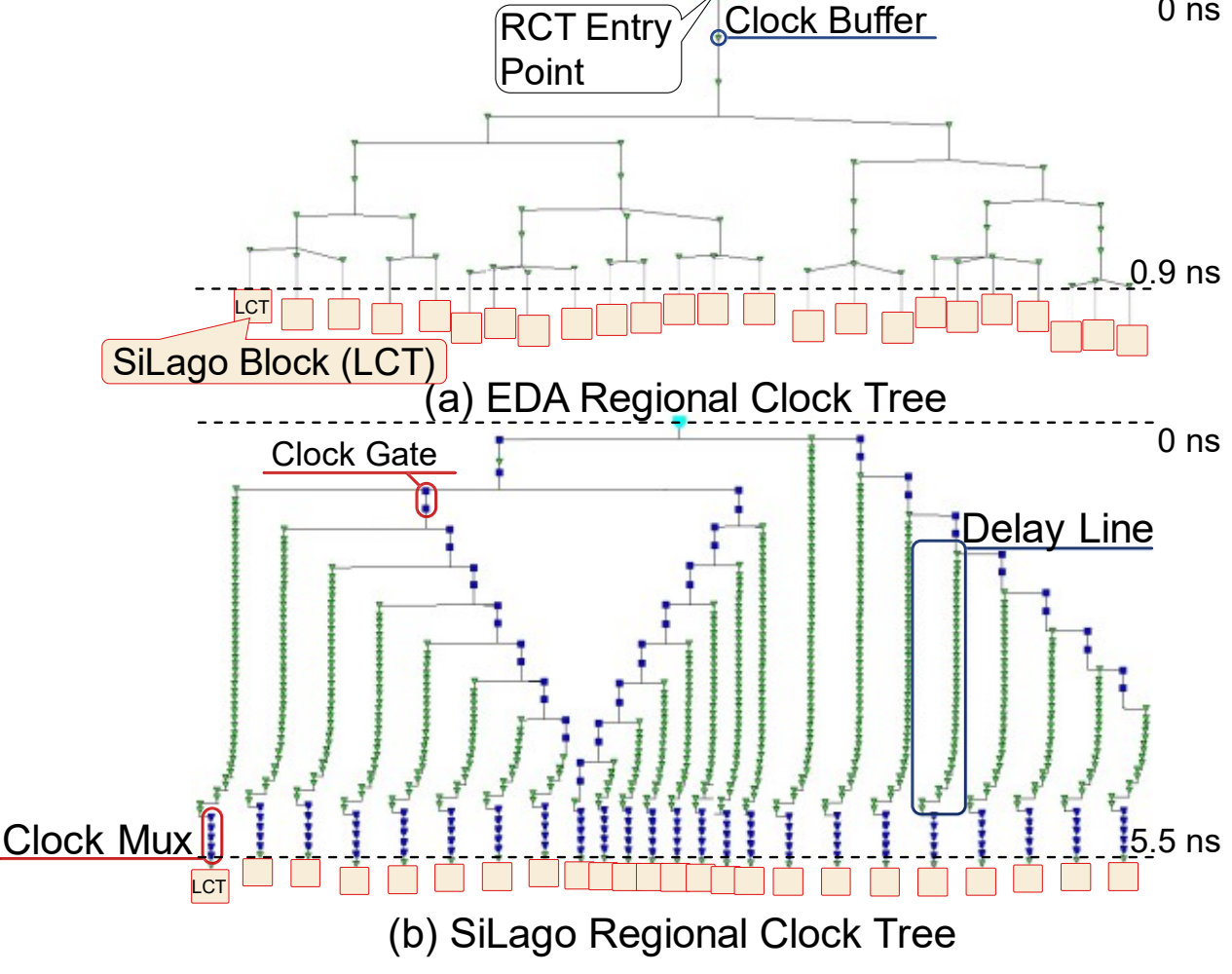

Figure 15. EDA generated RCT vs SiLago RCT.

## VI. State of the Art

The CTS (Clock Tree Synthesis) has been researched since the earliest days of VLSI. The CTS research's main objective has been to optimise the clock tree's cost metrics, i.e., smaller switching capacitance, minimising skew, maintaining slew rate, etc. [15], [16]. These approaches are based on sophisticated heuristic algorithms. Most of these methods are based on the van Ginneken dynamic programming algorithm for buffer insertion and sizing, and the delay model used is the Elmore delay model [17]. An excellent survey of CTS techniques is presented in [18]. In [15], [19], [20] the authors use a post-clock-tree-synthesis optimization. The proposed methods optimise the clock tree by altering an already synthesised tree. In contrast, our goal is to balance the clock's arrival time at the local clock trees' source points.

The key difference between the research proposed in this paper and the ad hoc CTS research reviewed here is that in the synchoros VLSI design style, composing a design in terms of SiLago blocks does not involve creating new wires of any type, including the clock. In the state-of-the-art, clock wires and buffers are synthesised as part of physical synthesis as a post logic synthesis step. In other respects, the RCT generation method proposed in this paper does not compete directly with the research reviewed above; the two methods are complementary. We suggest an alternative to the clock tree generation at a higher level, which has relatively insignificant capacitance overhead but significantly lowers the engineering cost and improves predictability.

CTS research has also focussed on regular topologies of clock trees like H-tree and Mesh clock structures [17], [21]. The regularity of these structures makes them *seem* like a good match for the synchoros VLSI design style. These structures do not fulfil the requirements of creating RCT by abutment. H-Tree is a hierarchical structure, and its depth would depend on the size of the region instance. For this reason, absorbing an H-Tree as RCT fragments and creating an arbitrary H-Tree by abutment that depends on the size and shape of the region instance is not feasible. Mesh-based clock tree poses a challenge in terms of the predictability requirement. Since the clock tree mesh creates an equipotential surface, it creates cyclic graphs that are impossible to analyse with STA [17], [21]. This is an even bigger problem for the RCT by abutment scheme because it requires the iron-clad timing models to predict the RCT properties with sufficient accuracy. The standardised entry points and propagation path of the SiLago RCT make it possible to have a trustworthy timing model.

Like synchoros VLSI designs, FPGAs are regular structures; it is natural to compare their clock tree schemes. The fundamental difference is that the clock tree routing is pre-designed for each FPGA. The FPGA wires and LUTs can be configured differently, but the silicon will not change. In contrast, the higher abstraction synthesis will create region instances of different sizes, aspect ratios, and entry points depending on the functionality and constraints. This will result in different RCT dimensions and topologies. All the wires in synchoros VLSI design are regular and structured but can be connected and organised in various ways to create endless variations of structures. Further, FPGAs do require an STA as a post-synthesis step. In synchoros VLSI design, the regions are pre-characterised to work up to a clock speed, as long as the RCT infrastructure can deliver the clock to the LCT entry points inside a known skew margin. This is why the synchoros VLSI design requires no physical level STA. The region instances of arbitrary size and domain-specific functionality are guaranteed to be correct-by-construction.

Some methods have been proposed for post-fabrication clock deskewing [22], [23]. These methods use configurable buffers and delay lines to correct the skew after fabrication and are still designed ad-hoc for each design. Our method's innovation is the structured design of the RCT that makes it predictable and not the use of the delay line. The structured and predictable design enables higher-level synthesis, as argued in [2].

## VII. Conclusion and Future Work

We have presented a regional clock tree generation method based on abutting fragments of RCT that are absorbed SiLago blocks as a one-time effort. The generated RCT is configurable, predictable and can be analyzed from higher abstractions. Additionally, we analyzed the tap selection problem and proposed a heuristics set to shrink the search space. The SiLago RCT was compared with an equivalent EDA generated RCT, and we show that in our experimental platform, the design metrics are comparable.

We are working on several enhancements to the RCT method and the overall synchoros VLSI design framework. The programmable delay line is constructed from qualified standard cells, and its atomic delay decides the resolution to which we can minimise the skew. We are working on a more advanced programmable delay line that will allow a finer adjustment of delay that would be needed for higher frequencies. In this improved delay line, the number of taps would increase logarithmically with the delay. A delay line will be constructed with weighted positional taps like a fixed-point number. We are also working on making the delay of each tap in the delay line adaptive [24], [25] to make the design more robust and enable dynamic voltage frequency scaling.

## Acknowledgment

This work was supported by Vinnova, Sweden, as part of the CREST II project and is now supported by ECSEL JU / Vinnova, Sweden funded project StorAIge.

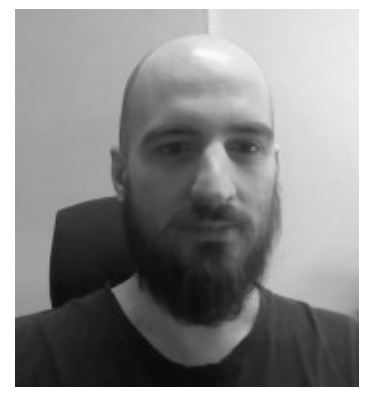

**Dimitrios Stathis** received his Diploma in Electr. & Comp. Engineering and a M.Sc. in Microelectronics and Computer Systems, both from the Democritus University of Thrace, Xanthi, Greece in 2013 and 2016 respectively. He received his second M.Sc. degree in Embedded Systems from KTH, Stockholm, Sweden in 2017. He is currently working towards the PhD degree at the KTH, Stockholm, Sweden. His research interests include Computer Architectures and ASIC design for multimedia applications and Artificial Neural Networks.

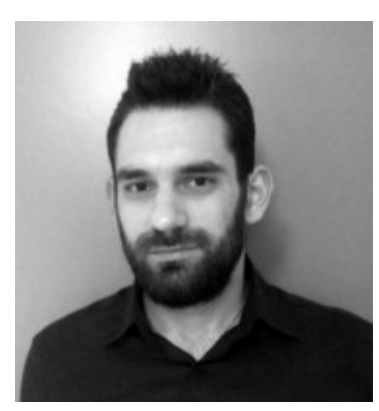

**Panagiotis Chaourani** received the B.Sc. degree in Physics and the M.Sc. degree in Electronics, both from the Aristotle University of Thessaloniki, Thessaloniki, Greece in 2011 and 2014 respectively. He received his PhD degree at the KTH, Stockholm, Sweden in 2019. His research interests include IC design, CAD tools and monolithic 3-D integration.

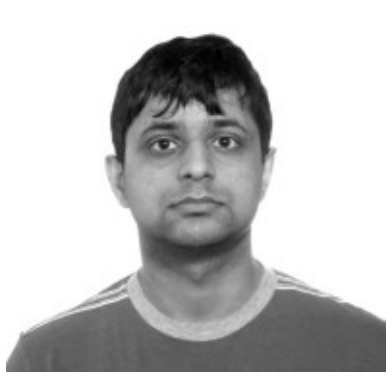

**Dr. Syed M. A. H. Jafri** is a senior system architect working at Ericsson. Before that, he worked as a post-doc researcher at the Royal Institute of Technology (KTH). He received his B.Sc. degree in 2005 from National University of Sciences and Technology in Rawalpindi, Pakistan. From 2005 to 2007 he was with Siemens, Pakistan. In 2009 he received MSc in system on chip design from Royal Institute of Technology (KTH), Sweden. He did PhD from University of Turku, Finland in 2015. He has 40+ peer-reviewed journals/conference publications.

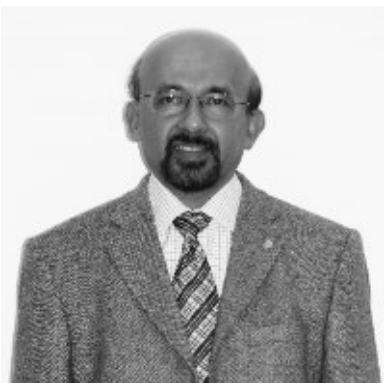

**Ahmed Hemani** is Professor in the Electronics Dept. at School of EECS, KTH, Stockholm, Sweden since 2006. His doctoral thesis on HLS was the basis for one of the first commercial HLS product from CADENCE. Later he has contributed to latency insensitive design style with research on GALS and GRLS. He also pioneered the concept of NOCs. His current research interests focus on massively parallel architecture and design methods for them to achieve ASIC comparable performance. He has introduced the concept of synchoricity and driven the development of the SiLago as an experimental synchoros VLSI design platform. He is applying synchoros VLSI design to artificial neural networks, biologically plausible models of brain and complex multi-media and telecom applications.